\documentclass[two column,prb,showpacs,multicol,amsmath,amssymb]{revtex4}
\usepackage[dvips]{graphicx}
\usepackage{graphicx}
\usepackage{dcolumn}
\usepackage{bm}
\usepackage{graphics}
\usepackage{soul}

\usepackage{epsfig,color}

\newcommand{\be}{\begin{equation}}
\newcommand{\ee}{\end{equation}}
\newcommand{\bea}{\begin{eqnarray}}
\newcommand{\eea}{\end{eqnarray}}

\begin{document}
\title{Zero-temperature study of a tetrameric spin-$1/2$ chain in a transverse magnetic field }
\author{J. Vahedi$^{1,2} \footnote{email: javahedi@gmail.com\\ Tel: (+98)9111554504\\Fax: (+98)15133251506}$, M. Shabani Arbousara$^3$, S. Mahdavifar$^3$}
\address{$^1$Department of Physics, Sari Branch, Islamic Azad University, Sari, Iran.\\
$^2$ Center for Theoretical Physics of Complex Systems, Institute for Basic Science (IBS), Daejeon, Korea.\\
 $^3$Department of Physics, University of Guilan, 41335-1914, Rasht, Iran.}
\date{\today}

\begin{abstract}
We consider an alternating Heisenberg spin-$1/2$ antiferromagnetic-ferromagnetic ($AF-F$) chain with the space modulated dominant antiferromagnetic exchange and anisotropic ferromagnetic coupling (tetrameric spin-$1/2$ chain). The zero-temperature effect of a symmetry breaking transverse magnetic field on the model is studied numerically. It is found that the anisotropy effect on the ferromagnetic coupling induces two new gapped phases. We identified their orderings as a kind of the stripe-antiferromagnetic phase. As a result, the magnetic phase diagram of the tetrameric chain shows five gapped quantum phases and the system is characterized by four critical fields which mark quantum phase transitions in the ground state of the system with the changing transverse magnetic field. We have also exploited the well known bipartite entanglement (name as concurrence) and global entanglement tools to verify the occurrence of quantum phase transitions and the corresponding critical points.
\end{abstract}
\pacs{ 75.10.Pq, 75.10.Hk}
\maketitle
\section{Introduction }\label{sec-I }

At zero temperature, quantum fluctuations play the
dominant role in determining the ground state properties of the physical system. The quantum fluctuations cause a fundamental change in the state of a system which is known as the quantum phase transition\cite{Sachdev01}. The one-dimensional (1D) bond alternating Heisenberg spin-$1/2$ models which are obtained by a space modulation in the exchange couplings represent one particular subclass of low-dimensional quantum magnets which pose interesting theoretical\cite{Takada92,Hida92,Kohmoto92,Yamanaka93,Sakai95,Uhrig96,Barnes99,Bocquet00,Yamamoto05,Zheng06,
Mahdavifar08,Abouie08,Guang09,Schmitt14,Liu145,Willenberg15, Tzeng16} and experimental \cite{16,17,18,19,20,21,22,23,24,s6,s8} problems. There is a spin-gap in the excitation spectrum of the  bond alternating spin-$1/2$ chains. The mentioned spin-gap causes a plateau shape  in the curve of the response functions specially the magnetization.

The study of the induced effects of the space modulation  on  the  exchange  couplings  has attracted much interest in recent years\cite{j1,j2,j3,j4,j5,j6,j7,j8,j9,j10,j11,j12}. It is known that the magnetization curve of a trimerized  Ferromagnetic-Ferromagnetic-Antiferromagnetic Heisenberg spin-1/2 chain shows a plateau at $1/3$ of the saturation\cite{j1,j2,j3,j7} but in a Ferromagnetic-Antiferromagnetic-Antiferromagneti  chain a mid-plateau is reported\cite{j6,j7}. Another kind of the alternating chains is known as the tetrameric chain\cite{j4,j5,j6,j8,j9}. In this model a mid-plateau  is also appeared  in  the  magnetization  curve by applying an external magnetic field.

\begin{figure}[t]
\includegraphics[width=0.95\columnwidth]{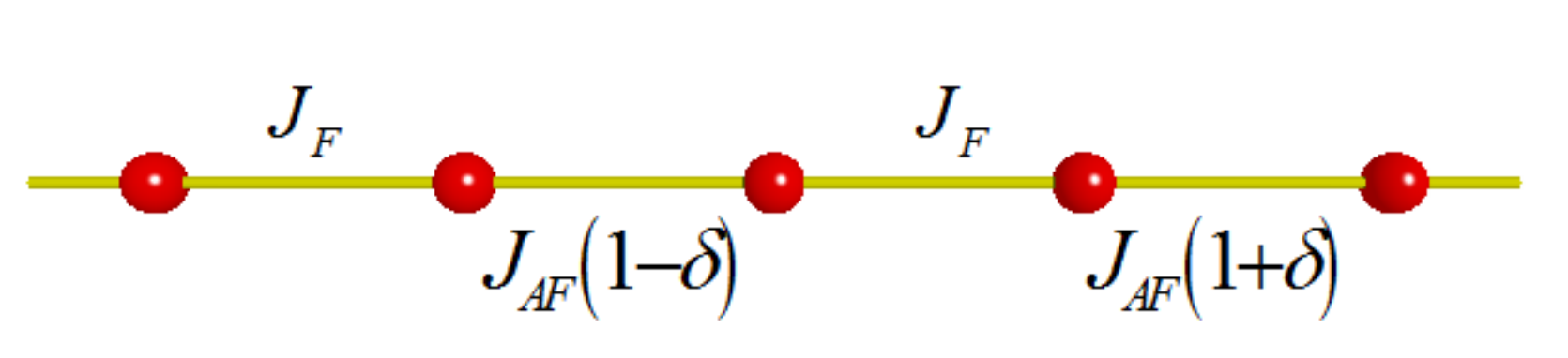}
\caption{(Color online.) The schematic picture of alternating tetrameric spin-1/2 chain.}
\label{dimer}
\end{figure}

Here, we continue the study of the zero temperature physics of the tetrameric spin-$1/2$ Heisenberg chains (see Fig.~\ref{dimer}). The tetrameric model is defined as an alternating Heisenberg $AF-F$ chain with the space-modulated antiferromagnetic exchange\cite{j9}. The Hamiltonian of the model is written as
\begin{eqnarray}
\emph{H}&=&-J_{F}\sum_{j=1}^{N/2}\Big[S_{2j}^{x}S_{2j+1}^{x}+S_{2j}^{y}S_{2j+1}^{y}+\Delta S_{2j}^{z}S_{2j+1}^{z}\Big]\nonumber\\
&+&J_{AF}\sum_{j=1}^{N/2}\Big[1+(-1)^{j}\delta \Big]\textbf{S}_{2j-1}\cdot\textbf{S}_{2j}\nonumber\\
&-&h\sum_{j=1}^{N}S_{j}^{x},
\label{e1}
\end{eqnarray}
where $S_{j}$ is the spin-$1/2$ operator on the $j$-th site. $J_{F}$ and $J^{\pm}_{AF}=J_{AF}(1\pm\delta)$ denote the ferromagnetic and antiferromagnetic couplings respectively, $h$ is the uniform transverse magnetic field and $\Delta$ denotes the anisotropy parameter. It is clear that in the unite cell, there are four different links, two equal ferromagnetic and two non-equal antiferromegnetic links.

 For $J_{AF}=0$ and in the absence of the magnetic field, the spins on odd links form local triplets and the model reduces to the spin-1 chain. As soon as the magnetic field is applies all spins on odd links will be aligned along the magnetic field. On the other hand, in the absence of the space modulation, $\delta=0$, the model reduces to the well-known alternating Heisenberg spin-1/2 chains in a transverse magnetic field\cite{Mahdavifar08}. Two Ising-type quantum phase transitions in presence of the transverse magnetic field have been identified. A gapped phase named stripe-antiferromagnetic has been found in the ground state phase diagram. 

\begin{figure}[t]
\includegraphics[width=1\columnwidth]{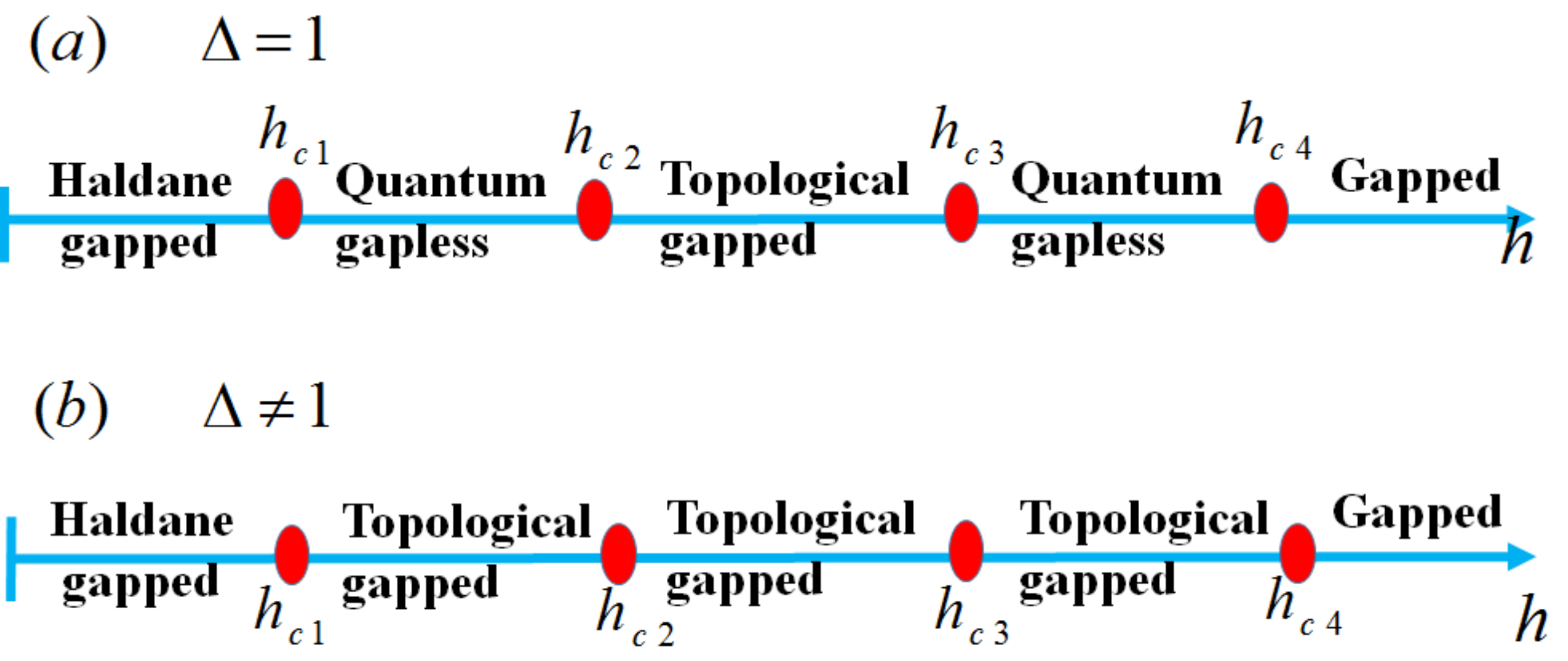}
\caption{(Color online) The phase diagram of the alternating tetrameric spin-1/2 chain in a transverse magnetic field: (a) Isotropic case, $\Delta=1$, (b) Anisotropic case, $\Delta\neq1$. Important difference is the replacing of the quantum gapless phases of the isotropic case with topological gapped phases\cite{j13} in the anisotropic case.}
\label{Fig0}
\end{figure}

In presence of the space modulation, $\delta \neq 0$, the isotropic model ($\Delta=1$) is known\cite{j9} very well (see Fig.~\ref{Fig0} (a)). Four Ising-type quantum phase transitions happen by applying the magnetic field . In principle, compared with the bond alternating model, the space modulation in the isotropic case, induces a new gaped phase in the ground state phase diagram. By opening this new gap, a magnetization mid-plateau appears. 
\par
In this paper we study the effect of a transverse magnetic field on the ground state phase diagram of the model. First, by assuming that the antiferromagnetic couplings are dominant we show that the model can be regarded as an $XYZ$ chain in the mutual effect of the longitudinal and staggered magnetic fields. Then, to explore the nature of the spectrum and the phase transition, we used the Lanczos method to numerically diagonalize finite chains. Using the exact diagonalization results, we calculate the gap, the magnetization, the string order parameter, and different spin correlation functions versus the transverse magnetic field. Based on the numerical results, we show that five gapped phases exist in the ground state phase diagram (see Fig.~\ref{Fig0} (b)).  Finally, we study the quantum correlations as the entanglement and the global entanglement.
\par
In the next section, we briefly discuss the model in the strong antiferromagnetic coupling and map the model to an effective $XYZ$ model. In Sec. III we present numerical results on the ground state phase diagram of the system. In Sec. IV the results of the entanglement study are presented. Finally, we conclude and summarize our results in Sec. V.

\section{EFFECTIVE HAMILTONIAN }\label{sec-II }

In the considered limiting case of the strong antiferromagnetic coupling $J_{AF}>>J_{F}$ and strong magnetic field $h\simeq J_{AF}$, one can use standard procedure\cite{Mila98} to map the model onto an effective spin chain Hamiltonian, which allows us to outline the symmetry aspect of the problem under consideration. Let us start from the limit of $J_{F}=0$, where at $h=0$ the system reduces to the set of noninteracting block of pairs of spins in the singlet state. At $J_{AF}>>J_{F}$, the system behaves as a nearly independent block of pair spins. Indeed an individual block of pair spins is in the singlet $|S\rangle= \frac{1}{\sqrt{2}}[|\uparrow\downarrow\rangle - |\downarrow\uparrow\rangle]$ or one of the  triplet states $|T_{1}\rangle=|\uparrow\uparrow\rangle$, $|T_{0}\rangle=\frac{1}{\sqrt{2}}[|\uparrow\downarrow\rangle + |\downarrow\uparrow\rangle]$ and $|T_{-1}\rangle=|\downarrow\downarrow\rangle$ with the corresponding energies which are obtained as
\begin{eqnarray}
E(S)&=&-\frac{3}{4} J_{AF},~~~~~~~~~~~~ E(T_{0})=\frac{1}{4} J_{AF}, \nonumber\\
E(T_{1})&=& \frac{1}{4} J_{AF}-h,~~~~~~~~ E(T_{-1})=\frac{1}{4} J_{AF}+h,
\label{e2}
\end{eqnarray}
respectively. By applying the magnetic field, the energy of the triplet state $E(T_{1})$, decreases and at $h=J_{AF}$ forms together with the singlet state a doublet of almost degenerate state. Therefor, the singlet $|S\rangle$ and the triplet $|T_{1}\rangle$ states construct  a new subspace for an effective spin $\tau=1/2$ system. One can project the original Hamiltonian Eq.~(\ref{e1}) on the new singlet-triplet subspace
\begin{eqnarray}
|\Downarrow\rangle\equiv|S\rangle=\frac{1}{\sqrt{2}}[|\uparrow\downarrow\rangle - |\downarrow\uparrow\rangle], ~~~~ |\Uparrow\rangle\equiv|T_{1}\rangle=|\uparrow\uparrow\rangle.\label{e3}
\end{eqnarray}
The relation between the real spin operator $\textbf{S}_{j}$ and the pseudo-spin operator ${\tau}$ in this restricted subspace can be easily derived to the first order and up to  a constant, we easily obtain the effective Hamiltonian as

\begin{eqnarray}
H^{eff}&=&-\frac{J_{F}}{2}\sum_{j=1,1}^{N/2}\Big[\Delta\tau_{j}^{x}\tau_{j+1}^{x}+\tau_{j}^{y}\tau_{j+1}^{y}+\frac{1}{2}\tau_{j}^{z}\tau_{j+1}^{z}\Big]\nonumber\\
&-&h_{0}^{eff}\sum_{j=1}^{N/2}\tau_{j}^{z}-h_{1}^{eff}\sum_{j=1}^{N/2}\tau_{j}^{z}(-1)^{j},
\label{e11}
\end{eqnarray}
where $h_{0}^{eff}=h-J_{AF}+\frac{J_{F}}{4}$ and $h_{1}^{eff}=\delta J_{AF}$. Note that in deriving
 $(4)$, we have used the rotation in the effective spin space which interchanges the $x$ and $z$ axes.
 At  $\Delta=1$, the effective Hamiltonian is nothing but the $XXZ$ Heisenberg chain in the presence of the uniform and staggered longitudinal magnetic fields\cite{Alcaraz95, Lou05,Mahdavifar07}. In addition, at $\Delta=\frac{1}{2}$, the effective model is known as the $XXZ$ model in transverse uniform and staggered magnetic fields\cite{Moradmard14}.  Away from the isotropic point, the effective Hamiltonian  describes the fully anisotropic ferromagnetic $XYZ$ chain in a space modulated magnetic field.
\section{numerical results} \label{secIII}
In this section, to explore the nature of the spectrum and the quantum phase transition, we used the Lanczos method to diagonalize numerically finite chains with lengths $N=12, 16, 20, 24$. To get the energies of the few lowest eigenstates we considered chains with periodic boundary conditions. We start our consideration by the anisotropic case, $\Delta\neq1$. First, we have computed the three lowest energy eigenvalues of chains with $J_{F}=1.0, J_{AF}=\frac{9}{2}, \delta=\frac{1}{9}$ and different values of the anisotropy parameter $\Delta=0.25, 0.5, 0.75$.
\begin{figure}[t]
\includegraphics[width=0.9\columnwidth]{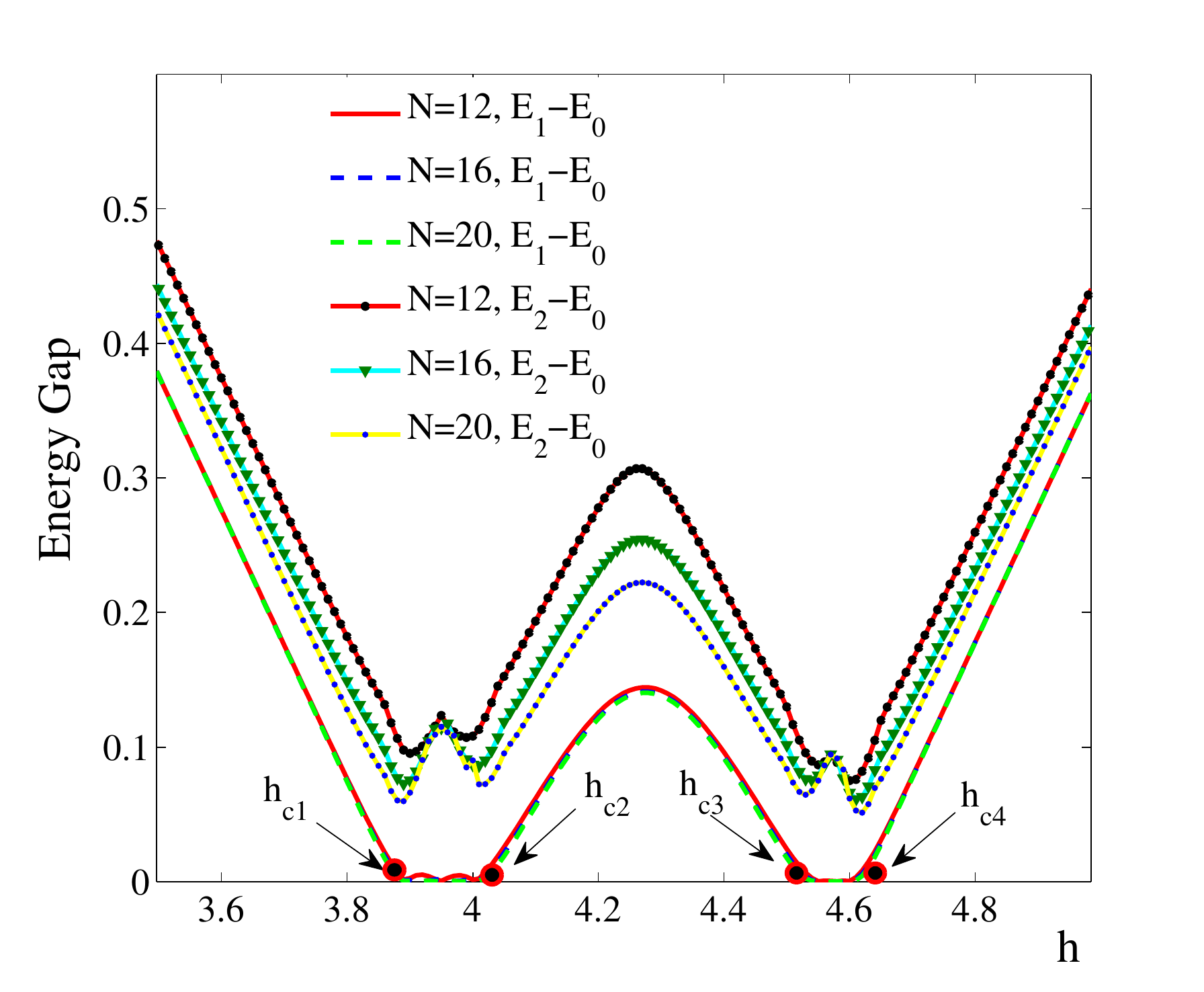}
\caption{(Color online). Difference between the energy of the two lowest levels as a function
of the magnetic field $h$, for chains with exchanges $J_{F}=1.0,J_{AF}=\frac{9}{2},\delta=\frac{1}{9}$
and lengths $N= 12, 16, 20$ and $\Delta=0.5$ .The three lines for $E_{1}-E_{0}$ are distinguishable on this scale.}
\label{gap}
\end{figure}
In Fig.~\ref{gap}, we have plotted results of these calculations for the anisotropy parameter $\Delta=0.5$. It can be seen that, in the absence of the magnetic field, the spectrum of the model is gapful. As soon as the magnetic field is applied, the energy of exited states decreases linearly and  vanishes ($N$ goes to infinity) at $h_{c_{1}}$, which is size independent. We define the energy gap as a difference between the second excited energy and the ground state energy. By more increasing the magnetic field, the energy gap which appears at $h>h_{c_{1}}$ first increases, then starts to decrease and again vanishes at $h_{c_{2}}$. With the continuing increase in the magnetic field, the similar incident repeated twice. In principle, there are five gapped phases in the ground state phase diagram of the system which are separated by four critical fields,
\begin{eqnarray}
h_{c_{1}}&=&3.86\pm0.01, \nonumber \\
h_{c_{2}}&=&4.01\pm0.01, \nonumber \\
h_{c_{3}}&=&4.53\pm0.01, \nonumber \\
h_{c_{4}}&=&4.61\pm0.01.
\label{e11-1}
\end{eqnarray}
It should be noted, that only in two gapped phases: $(I)~h_{c_{1}} < h < h_{c_{2}}, (II)~h_{c_{3}} < h < h_{c_{4}}$ the ground state energy is doubly degenerate. In the region $h>h_{c_{2}}$, the gap becomes proportional to $h$. Here, we should address that the mechanism of the opening gap is due to the alternation of a spin-1/2 chain in the region $h<h_{c_1}$, the anisotropy on the FM exchanges in the regions $h_{c_1} < h < h_{c_2}$,  $h_{c_3} < h < h_{c_4}$, the space-modulation on the AF exchanges in the region $h_{c_2} < h < h_{c_3}$, and  the magnetic field in the saturated region $h > h_{c_4}$.
\begin{figure}[t]
\includegraphics[width=0.9\columnwidth]{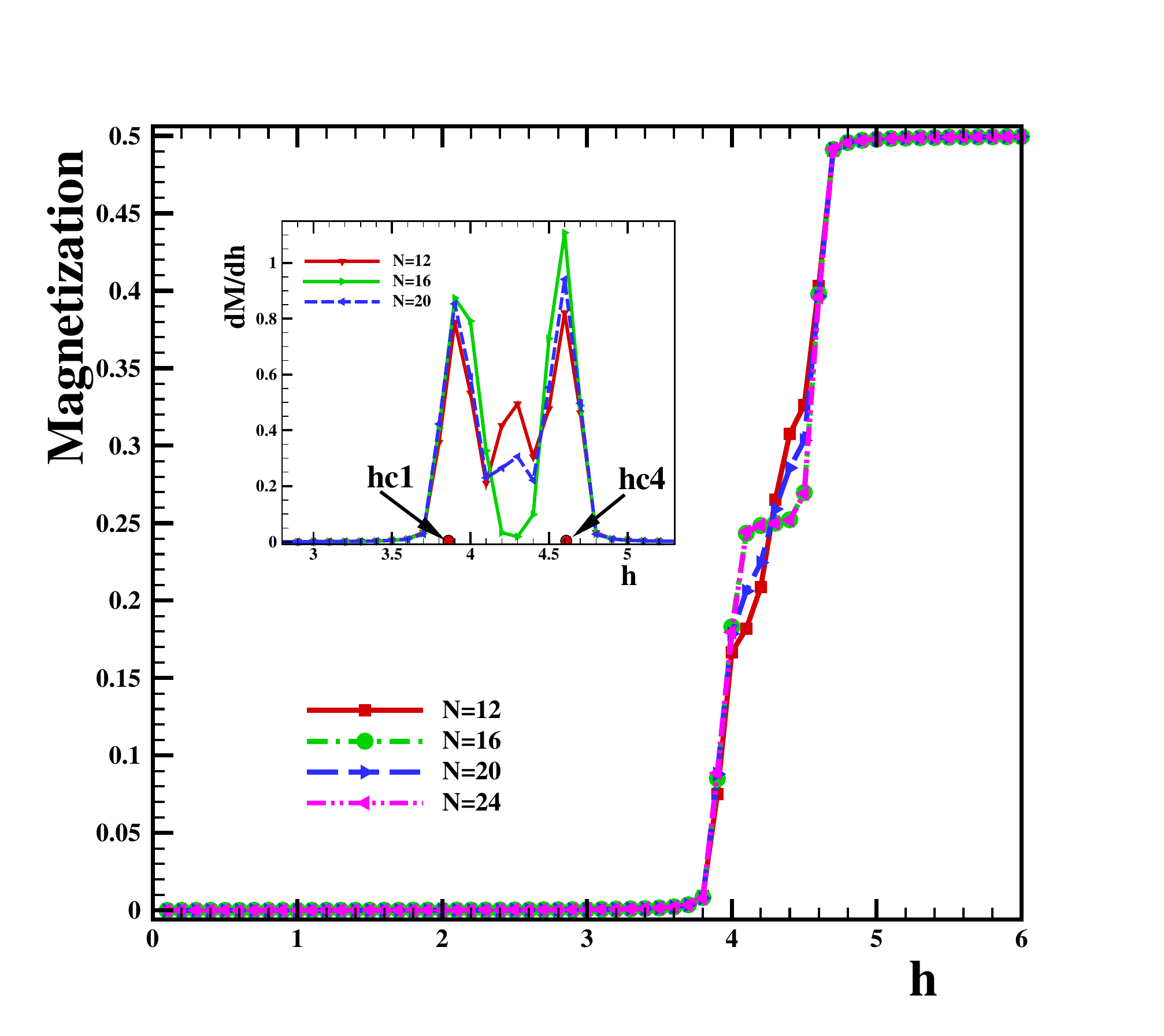}
\caption{(Color online.)  The magnetization along the magnetic field $M^{x}$ as function of applied field $h$, for chains with exchanges $J_{F}=1.0,J_{AF}=\frac{9}{2},\delta=\frac{1}{9}$ and lengths $N= 12, 16, 20, 24$ and $\Delta=0.5.$ The inset has shown the first derivation of the magnetization.}
\label{magnetization}
\end{figure}
\par
To recognize the different phases induced by the transverse magnetic field in the ground state phase diagram, we have implemented the Lanczos algorithm and calculated the lowest eignstates, the order parameter and various spin-spin correlation functions. A deep insight into the nature of the different phases can be obtained by studying the magnetization process. The magnetization along the transverse magnetic field axis is defined as
\begin{eqnarray}
M^{x}=\frac{1}{N} \sum_{n=1}^{N} \langle{Gs}|S_{n}^{x}|{Gs}\rangle,
\label{Neel}
\end{eqnarray}
where the notation $ \langle{Gs}|...|{Gs}\rangle$ represents the expectation value at the ground state. In Fig.~\ref{magnetization}, we have plotted $M^{x}$ as a function of the magnetic field $h$. For arriving at this plot we considered exchange parameters $J_{F}=1.0,J_{AF}=\frac{9}{2},\delta=\frac{1}{9}$ and anisotropy parameter $\Delta=0.5$.  The standard singlet and saturated ferromagnetic plateaus at $h<h_{c_{1}}$  and $h>h_{c_{2}}$ are observed. Due to the profound  effect of the quantum fluctuations, the transverse magnetization remains small but finite for $0<h<h_{c_{1}}$ and reaches zero at $h=0$. This behavior is in agreement with expectations, based on general statement  that in a gapped phase, the magnetization along the applied field appears only at a finite critical value of the magnetic field equal to the energy gap. By more increasing the magnetic field, magnetization increases for $h>h_{c_{1}}$ very fast.  However, in finite size systems we do not observe a sharp transition close to the saturation value, which happens at $h>h_{c_{4}}$. The values of the critical fields $h_{c_{1}}$ and $h_{c_{4}}$ depend on the anisotropy parameter $\Delta$. By increasing $\Delta$, the critical fields take larger values. Also, our numerical results show that the magnetizations along the directions perpendicular to the applied field, $M^{y}$ and $M^{z}$ remain zero. Critical fields can be also determined from the anomalies in the first derivation of some  physical functions such as the magnetization versus $h$. The inset of Fig.~\ref{magnetization} shows $\frac{dM}{dh}$ as a function of the magnetic field for different chain sizes $N=12,16,20$. As it is seen, the critical fields $h_{c_{1}}=3.86 \pm0.01$ and $h_{c_{4}}=4.61\pm0.01$ are determined from anomalies in the first derivation of the magnetization with respect to the $h$.
\begin{figure}[t]
\includegraphics[width=0.9\columnwidth]{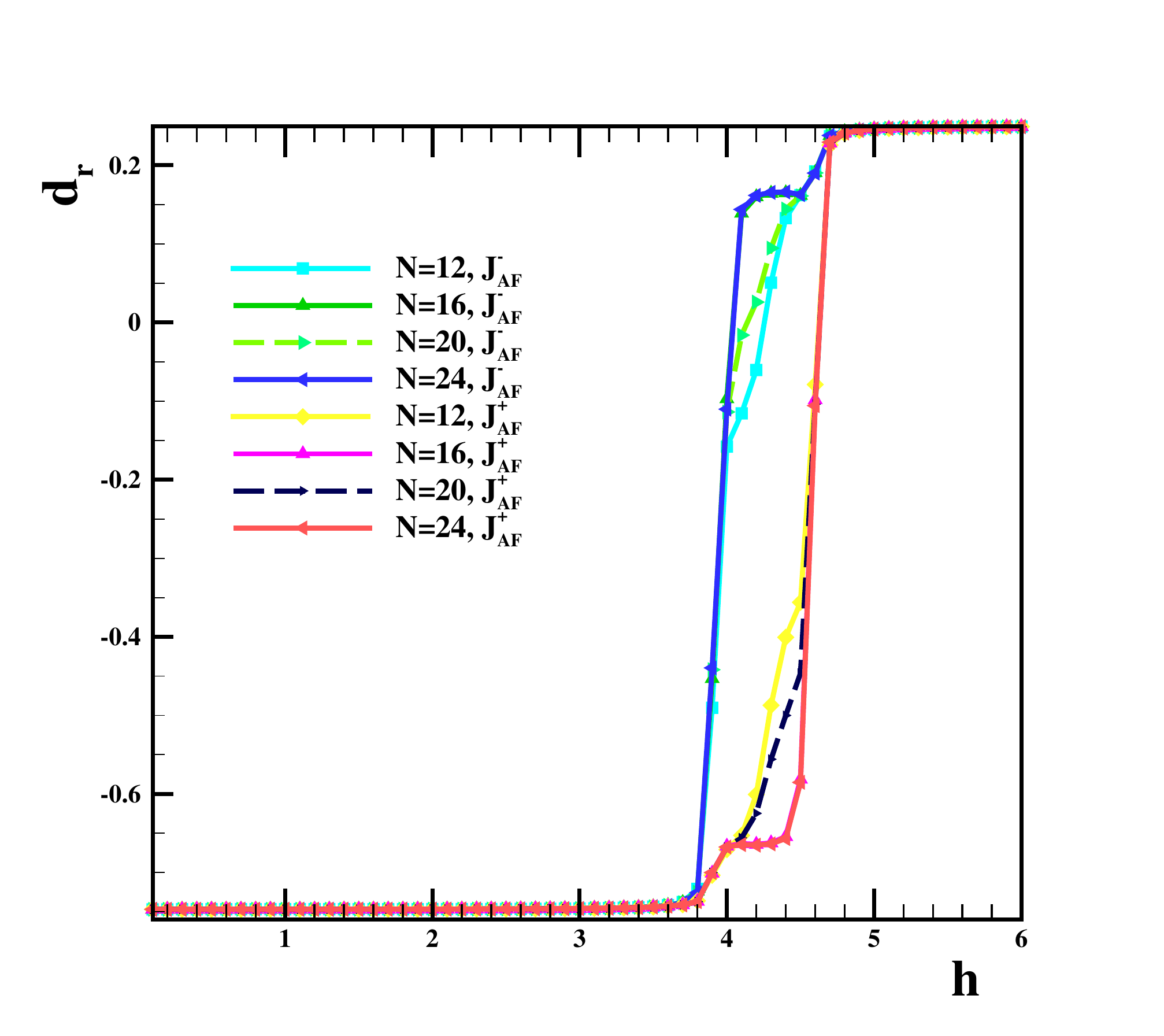}
\caption{(Color online). The $AF$-bond dimerization order parameter as a function of the applied field $h$, for chain with exchanges $J_{F}=1.0,J_{AF}=\frac{9}{2},\delta=\frac{1}{9}$ and lengths $N=12,16,20,24$.}
\label{phase}
\end{figure}
To get additional insight into the nature of different phases, we have also calculated the bond dimerization order parameter $d_r$. We define the antiferromagnetic bond dimerization  as
\begin{eqnarray}
d_{r}^{w}&=&\frac{4}{N}\sum_{j=1}^{N/2}\langle{Gs}|S_{2j-1}\cdot S_{2j}|{Gs}\rangle, \nonumber \\
d_{r}^{s}&=&\frac{4}{N}\sum_{j=2}^{N/2}\langle{Gs}|S_{2j-1}\cdot S_{2j}|{Gs}\rangle,
\label{dimer-w}
\end{eqnarray}
taking sum over antiferromagnetic weak (odd values of $j$) or strong (even values of $j$) bond, respectively. In Fig.~\ref{phase} we have plotted the $d_{r}^{w}$ and $d_{r}^{s}$ as a function of the magnetic field $h$, for chains with exchanges $J_{F}=1.0, J_{AF}=\frac{9}{2}, \delta=\frac{1}{9}$ and $\Delta=0.5$.  At the first glance, it is seen that in the region $h<h_{c_{1}}$ pair of spins on all antiferromagnetic bonds are in the singlet state with  $d_{r}^{w}=d_{r}^{s}\simeq\-0.75$, while at $h>h_{c_{4}}$, all spins  are aligned and $d_{r}^{w}=d_{r}^{s}\simeq\frac{1}{4}$. Numerical results in the intermediate region of the magnetic field, $h_{c_{1}}<h<h_{c_{4}}$, give us ability to trace the mechanism of singlet pair melting with respect to the field. As soon as the magnetic field increase from $h_{c_{1}}$, all spin singlet pairs start to melt simultaneously.  With further increase of $h$, melting of weak antiferromagnetic bonds gets more intensive, however at $h_{c_{2}}<h<h_{c_{3}}$ the process of melting are suppressed. As it is seen in Fig.~\ref{phase} weak antiferromagnetic bonds are suppressed, however their dimerization is far from the saturation value, while the strong  antiferromagnetic bonds  still manifest strong singlet features. In the region $h>h_{c_{3}}$, antiferromagnetic bonds starts to melt more intensively while weak antiferromagnetic bonds increase slowly. Finally at $h>h_{c_{4}}$ both subsystems of antiferromagnetic bonds achieve an identical almost fully polarized state.
\begin{figure}[t]
\includegraphics[width=0.9\columnwidth]{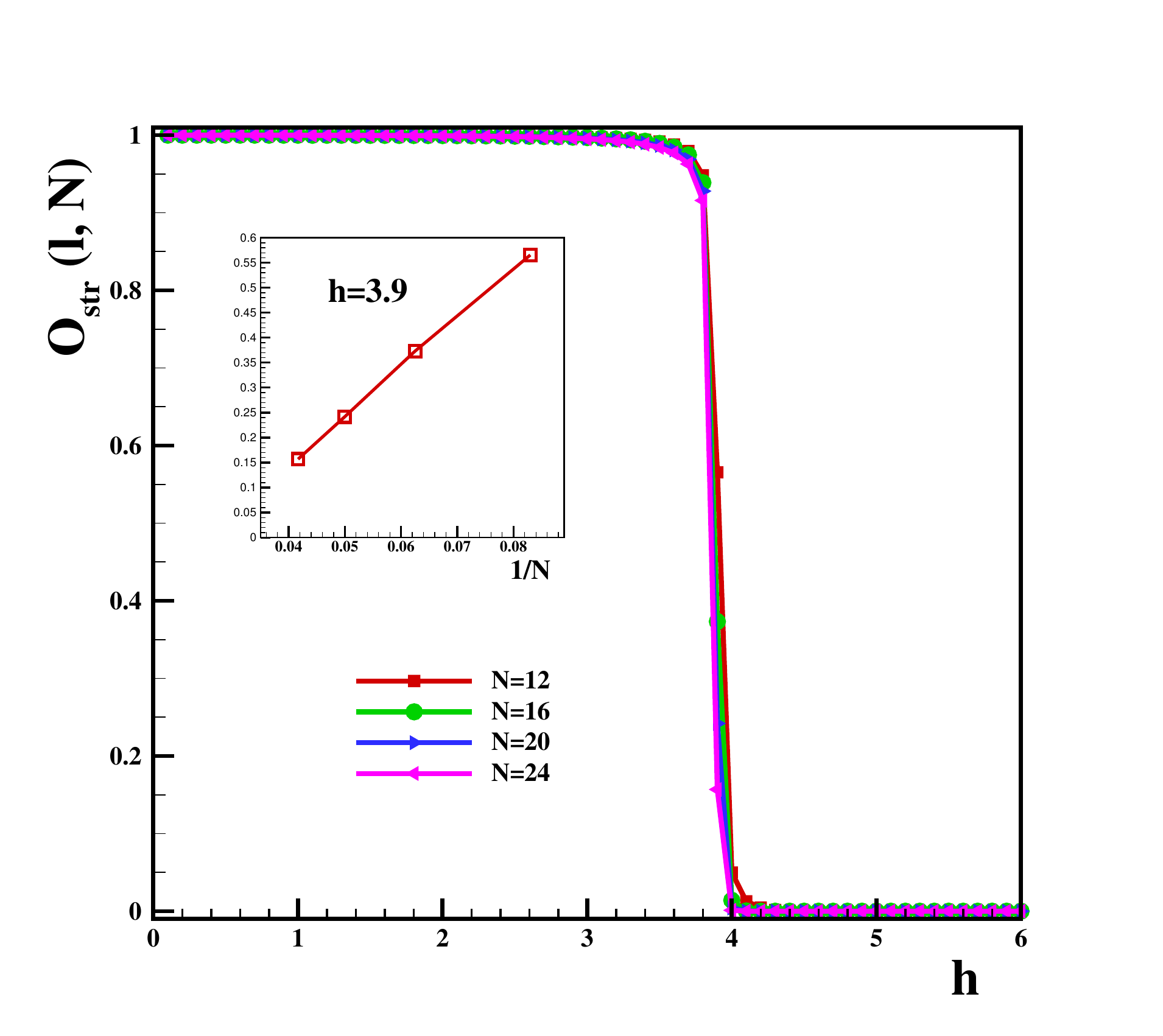}
\caption{(Color online). The string correlation function $\textbf{ O}_{str}(l,N)$ as function of the transverse magnetic field $h$ for different chain lengths $N=12,16,20,24$ with exchanges $J_{F}=1.0,J_{AF}=\frac{9}{2},\delta=\frac{1}{9}$ and $\Delta=0.5$.The inset show the string order parameter $\textbf{O}_{str}(l,N)$ as function of the $1/N$ for a value of magnetic field $h=111>h_{c_{1}}$.}
\label{string-1}
\end{figure}
\par
 By analyzing of the numerical results on the energy gap (Fig.~\ref{gap}), we found that the spectrum is gapful in the absence of the uniform transverse magnetic field which is one of the properties of the Haldane phase with long-range string order\cite{Haldane83}. The Haldane phase can be recognized from studying the string correlation function. The string correlation function in a chain of length $N$ defined only for odd $l$ as\cite{string}
\begin{eqnarray}
\textbf{O}_{str}(l,N)=-\langle{exp\{i\pi\sum_{k=2j+1}^{2j+1+l}}(S_{k}^{z})\}\rangle.
\label{string}
\end{eqnarray}
In particular, we have calculated the string correlation function for different finite chain lengths. (Since the present model has a $SU(2)$ symmetry in the absence of the magnetic field, we only consider the $Z$ component of the string correlation function.) In Fig.~\ref{string-1}, we have plotted $\textbf{O}_{str}(l,N)$ as a function of the magnetic field $h$ for different values of the chain lengths $N=12, 16, 20, 24$ with exchanges $J_{F}=1.0,J_{AF}=\frac{9}{2}, \delta=\frac{1}{9}$ and $\Delta=0.5$.
As can be seen from this figure, at $h<h_{c_{1}}$, $\textbf{O}_{str}(l,N)$ is close to its maximum value 1.0, therefore the tetrameric chain system is in the Haldane phase. The Haldane phase remains stable even in the presence of a transverse magnetic field less than $h_{c_{1}}$. In the inset of the Fig.~\ref{string-1}, we have also plotted the string order parameter, $\textbf{O}_{str}(l,N)$, as a function of $1/N$ for a value of magnetic field $h>h_{c_{1}}$. It is clear that by increasing the size of the system, the $\textbf{O}_{str}(l,N)$, converges to very small values close to zero, which shows that there is not the string  ordering at larger values of the transverse magnetic field $h>h_{c_{1}}$.
\section{ENTANGLEMENT STUDY} \label{secIV}
The quantum correlation, which known as entanglement is a purely quantum phenomenon with no classical counterpart. It is thought to hold the key to a deeper understanding of the theoretical aspects of quantum mechanics, in particular quantum spin models. It has been found that entanglement\cite{Wooters98} plays a crucial role in the low-temperature physics of many of these systems, particularly in their ground (zero temperature) state \cite{40,41,42,43}. It has been shown that the quantum phase transition (QPT) is accompanied by a profound change in the entanglement. Based on the different model, entanglement could peak, or show discontinuous manner, or exhibit diverging derivatives with scaling trend at the critical point\cite{Amico08}.
\subsection{CONCURRRENCE} \label{secA}
In this section we focus on  entanglement between two sites which is known as the concurrence. We compute the entanglement of two spins in different phases of the system, which allows us to verify the melting process. Concurrence is a measure of the bipartite entanglement which is defined as following\cite{Wooters98, Amico08}
\begin{eqnarray}
C_{lm}&=&2~max\{0, C_{lm}^{(1)}, C_{lm}^{(2)}\},\nonumber
\label{concurrence}
\end{eqnarray}
where
\begin{eqnarray}
C_{lm}^{(1)}&=& \sqrt{(g_{lm}^{xx}-g_{lm}^{yy})^{2}+(g_{lm}^{xy}+g_{lm}^{yx})^{2}} \nonumber \\
&-&\sqrt{(\frac{1}{4}-g_{lm}^{zz})^{2}-(\frac{M_{l}^{z}-M_{m}^{z}}{2})^{2}}\nonumber \\
C_{lm}^{(2)}&=& \sqrt{(g_{lm}^{xx}+g_{lm}^{yy})^{2}+(g_{lm}^{xy}-g_{lm}^{yx})^{2}} \nonumber \\
&-&\sqrt{(\frac{1}{4}+g_{lm}^{zz})^{2}-(\frac{M_{l}^{z}+M_{m}^{z}}{2})^{2}}.
\label{concurrence}
\end{eqnarray}
and $g_{lm}^{\alpha\beta}=\langle S_{l}^{\alpha} S_{m}^{\beta}\rangle$ is the correlation function between spins on sites $l$ and $m$.  In the region $h<h_{c_{1}}$, where the system is in the non-magnetic singlet phase, the correlations $g^{xx}$, $g^{yy}$, $g^{zz}$ are the same and take the value $-1/4$. Therefore the concurrence becomes one. On the other hand, for very large values of the magnetic field, $h>h_{c_{4}}$, the ground state of the system is written as
\begin{eqnarray}
|Gs\rangle=| \uparrow\uparrow\uparrow...\uparrow\rangle.
\label{dimer-GS}
\end{eqnarray}
In this saturated ferromagnetic phase, the value of correlations are $g^{xx}=g^{yy}=0,~ g^{zz}=1/4$ and $M^{z}=1/2$ and the concurrence vanishes. It can be seen that the numerical results, are in well agreement with what is expected. In principle our numerical experiment shows that in the absence of the magnetic field, spins are completely entangled.
.
\begin{figure}[t]
\includegraphics[width=0.9\columnwidth]{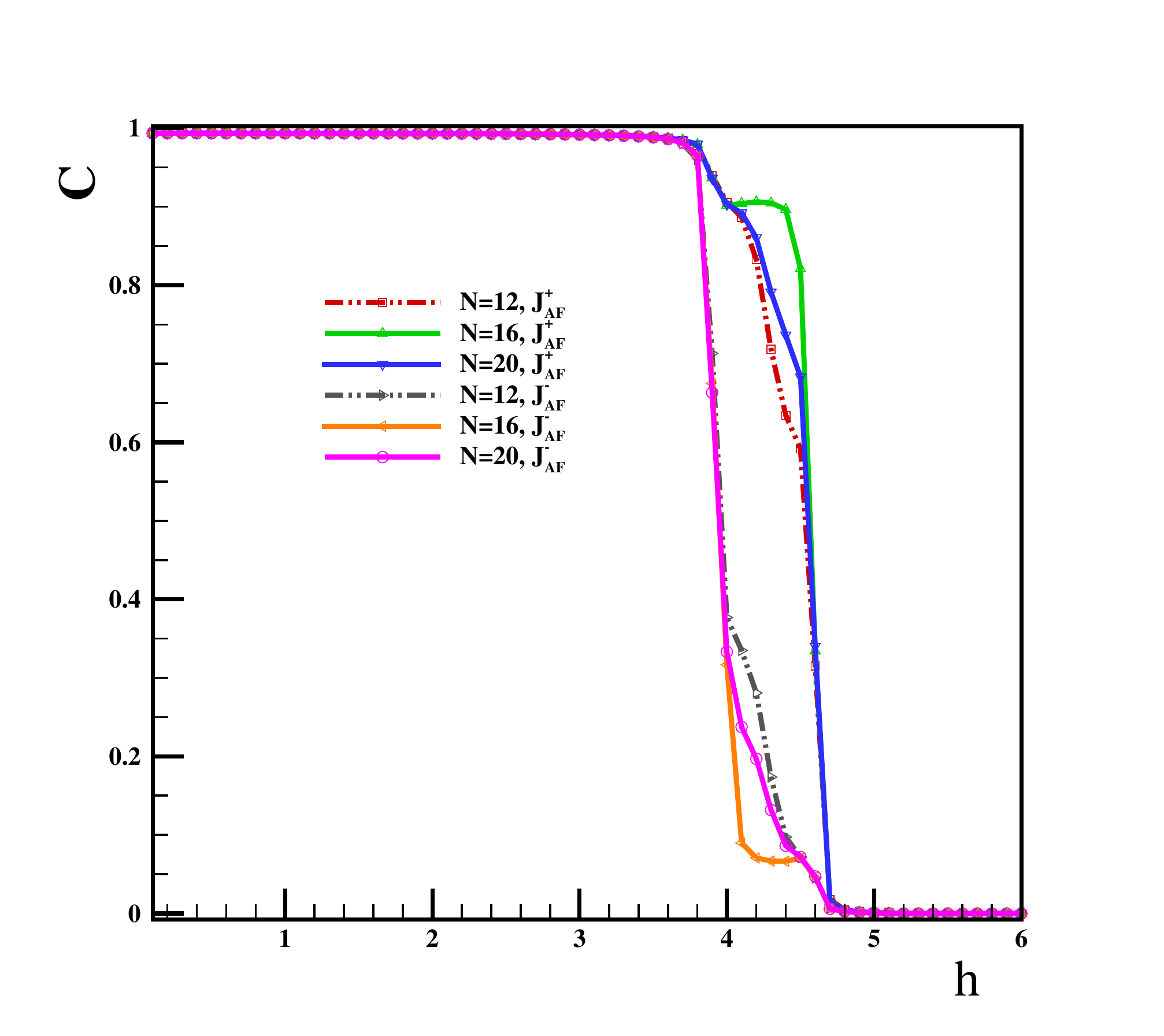}
\caption{(Color online). (a)The concurrence between two spins on strong $J_{AF}^+$ and weak  $J_{AF}^-$  links as a function of magnetic field versus applied magnetic field for chain with different lengths $N=12,16,20,24$,  exchange parameter $J_{F}=1.0,~J_{AF}=\frac{9}{2},~\delta=\frac{1}{9}$ and anisotropy parameter $\Delta=0.5$ .}
\label{concurrence}
\end{figure}
\begin{figure}[t]
\includegraphics[width=0.9\columnwidth]{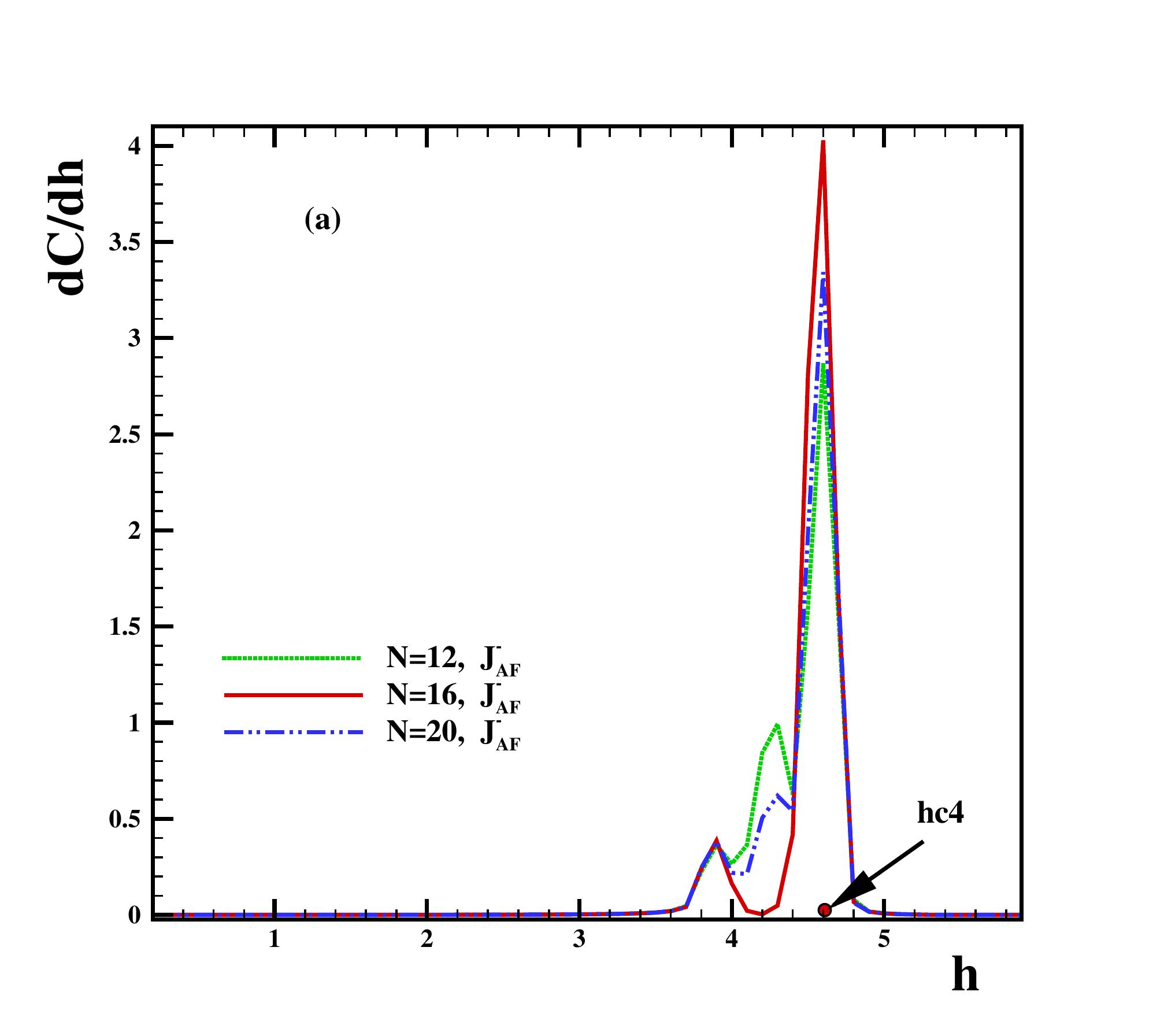}
\includegraphics[width=0.9\columnwidth]{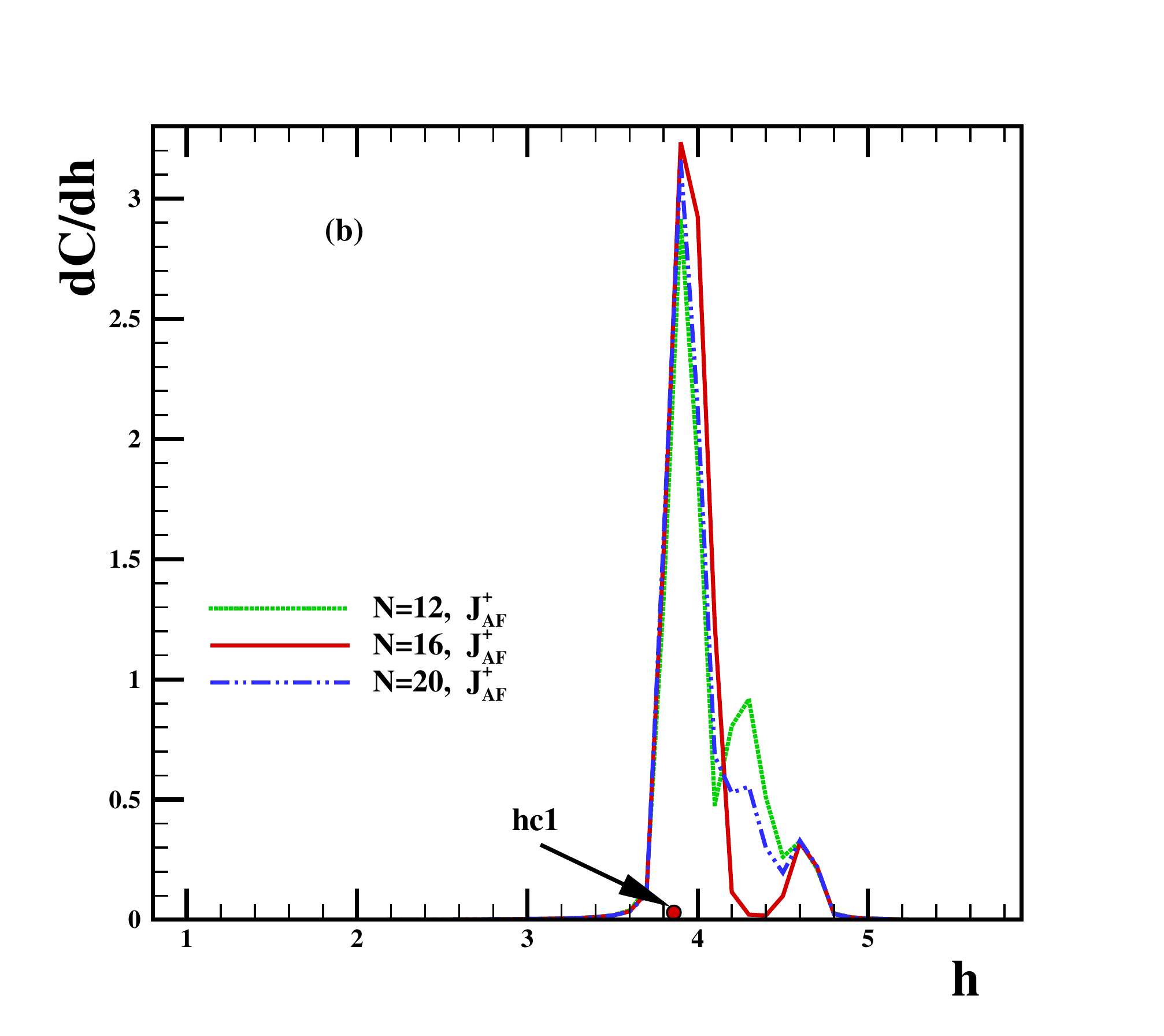}
\caption{(Color online).  The first derivation of the concurrence between two spins on (a) weak $AF_{1}$ and (b) strong $AF_{2}$ links respectively,  as a function of magnetic field for chain with exchange parameter $J_{F}=1.0,~J_{AF}=\frac{9}{2},~\delta=\frac{1}{9}$ and anisotropy parameter $\Delta=0.5$ .}
\label{dconcurrence}
\end{figure}
\par
We have plotted the entanglement of two spins which are located at the same strong or weak bond versus $h$ with different chain lengths $N=12, 16, 20, 24$ and exchange parameter $J_{F}=1.0,~J_{AF}=\frac{9}{2},~\delta=\frac{1}{9}$
and anisotropy parameter $\Delta=0.5$. Fig.~\ref{concurrence} shows the concurrence between two spins on strong $J_{AF}^+$ and weak $J_{AF}^-$ links as a function of magnetic field. It is clear that in the absence of the magnetic field, pair spins on weak and strong antiferromagnetic links are maximally entangled.
\par
Increasing of the magnetic field does not make change on the entanglement between pair spins on strong and weak antiferromagnetic bonds up to the first critical field $h_{c_{1}}$.  This behavior is in agreement that in the gapped singlet phase, the change in any physical function appears only at a critical value of the magnetic field equal to the gap. For $h>h_{c_{1}}$ the concurrence between spins on weak bonds $J_{AF}^-$ drops very rapidly when compared with strong exchanges $J_{AF}^+$. Indeed, the quantum correlations of the two spins with strong and weak antiferromagnetic exchanges decrease with increasing the magnetic field, but with the different intensity. In fact, such a intense quantum fluctuations cause to change in any physical function in the intermediate region. In the intermediate region the decreasing behavior of the concurrence continues up to the fourth critical field $h_{c_{4}}$, where takes the zero value. Finally, in the full-saturated ferromagnetic state, all of concurrence disappears and the entanglement of the state is exactly zero.
\par
Fig.~\ref{dconcurrence} shows the derivative of the concurrence with respect to the field, $\frac{dC}{dh}$, for weak $J_{AF}^-$ and strong  $J_{AF}^+$  links a function of the magnetic field for different chain size $N=12, 16, 20$, respectively. As can be clearly seen, the critical fields are $h_{c_{1}}=3.86 \pm0.01$ (bottom panel) and $h_{c_{4}}=4.61\pm0.01$ (top panel) from the anomalies in the first derivation of the concurrence curve as a function of $h$.
 \begin{figure}[t]
\includegraphics[width=0.9\columnwidth]{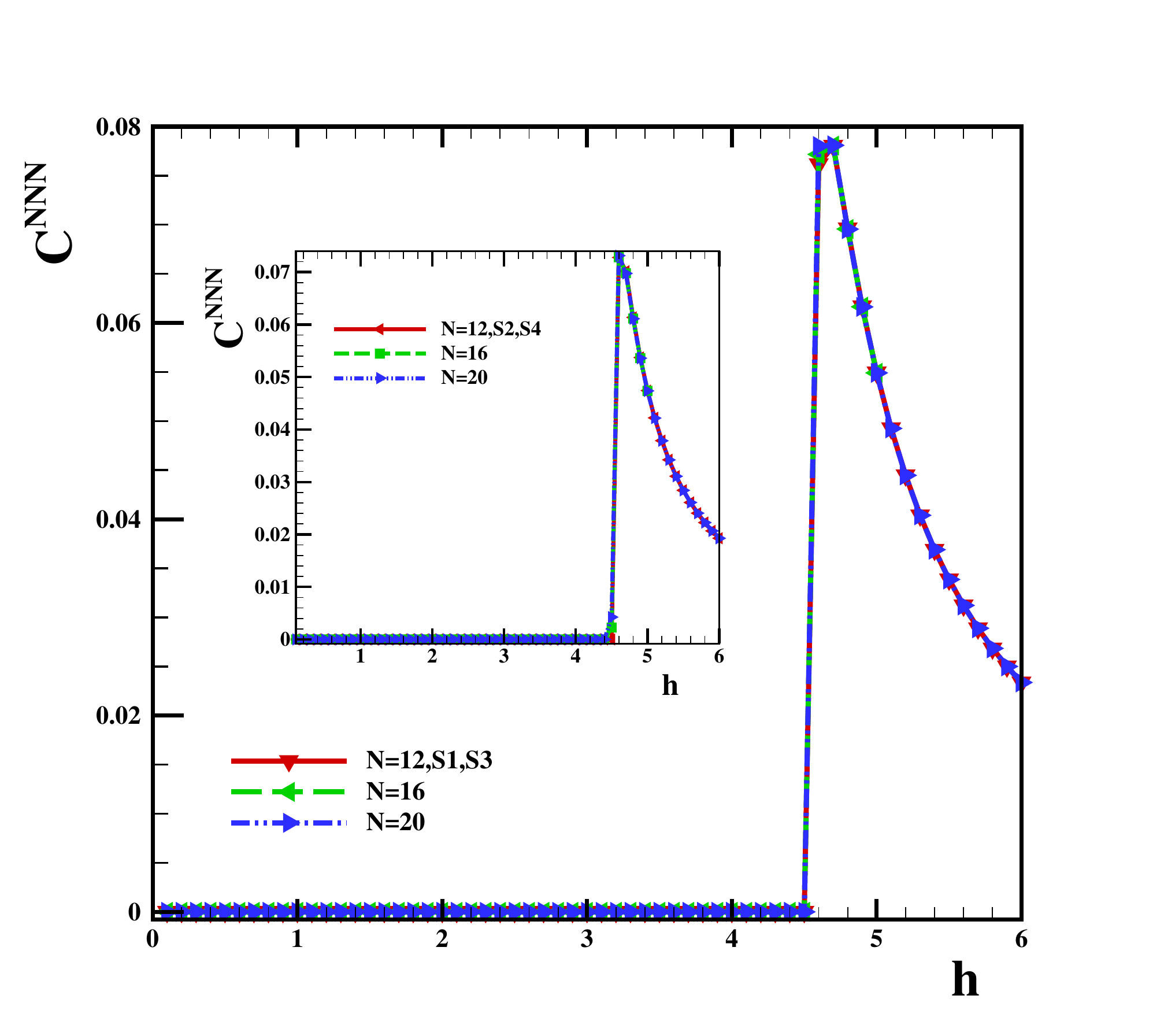}
\caption{(Color online). The concurrence between next nearest neighbors ($NNN$) spins as a function of $h$ for chains with exchanges $J_{F}=1.0,J_{AF}=\frac{9}{2},\delta=\frac{1}{9}$ and lengths $N= 12, 16, 20$ and $\Delta=0.5$.}
\label{concurrenceNNN}
\end{figure}
\par
To get more information about the entanglement between spins, we have also calculated the concurrence between next nearest neighbor ($NNN$) pair spins in the system. In Fig.~\ref{concurrenceNNN}, we have depicted the concurrence between $NNN$ pair spins on the odd and even site as a function of $h$. It is obvious that the $NNN$ pair spins are not entangled at $h=0$. It is also remarkable that from our numerical results we found that the concurrence between $NNN$ is zero up to $h_{c_{3}}$ . In the presence of the magnetic field the $NNN$ pair spins will be entangled only in the region $h_{c_{3}}<h<h_{c_{4}}$. It shows very sharp response to the applied magnetic filed at the critical value  $h_{c_{4}}$. Moreover, by increasing the magnetic filed the concurrence between $NNN$ pair spins starts to decrease monotonically.
 \begin{figure}[t]
\includegraphics[width=0.9\columnwidth]{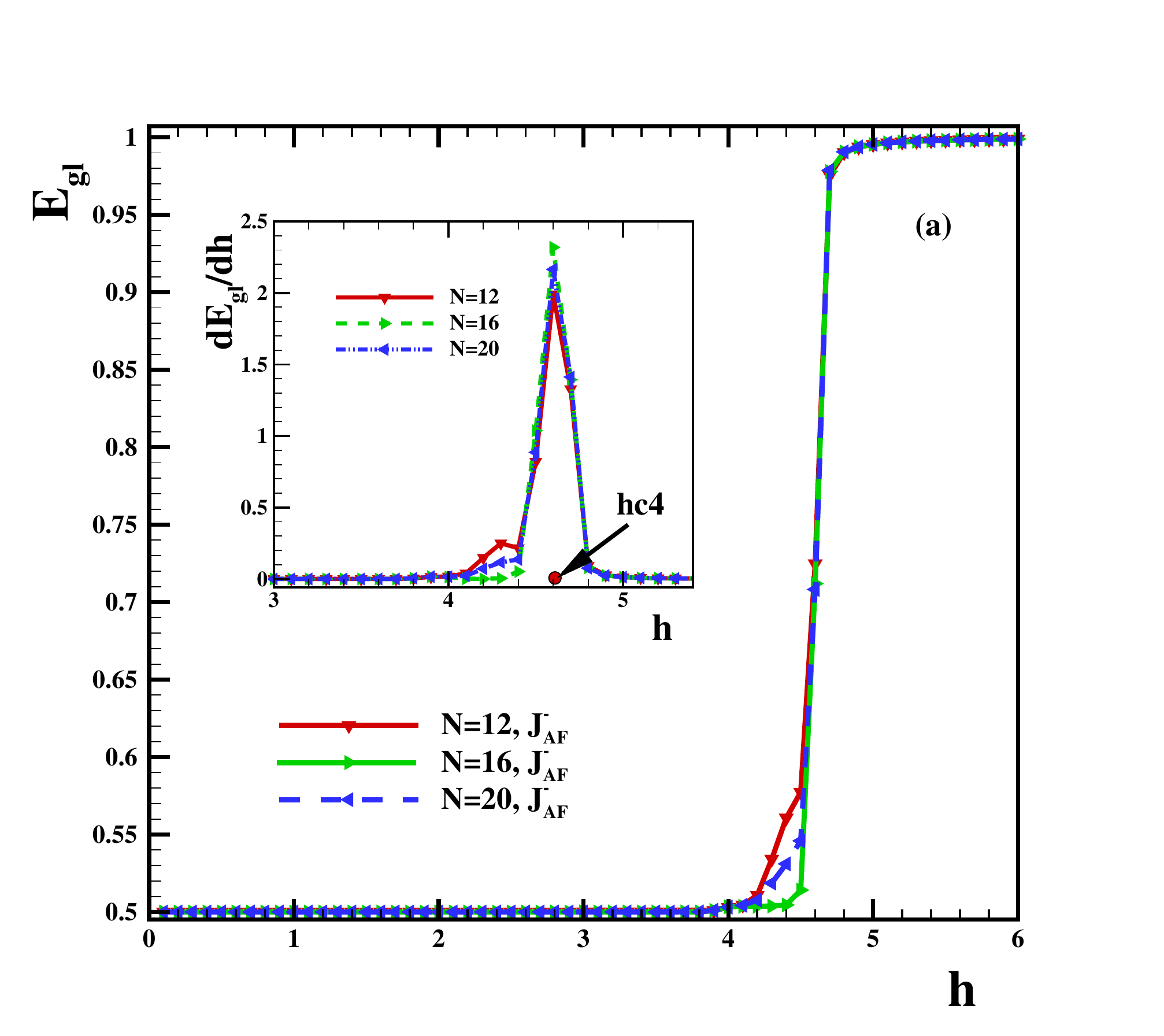}
\includegraphics[width=0.9\columnwidth]{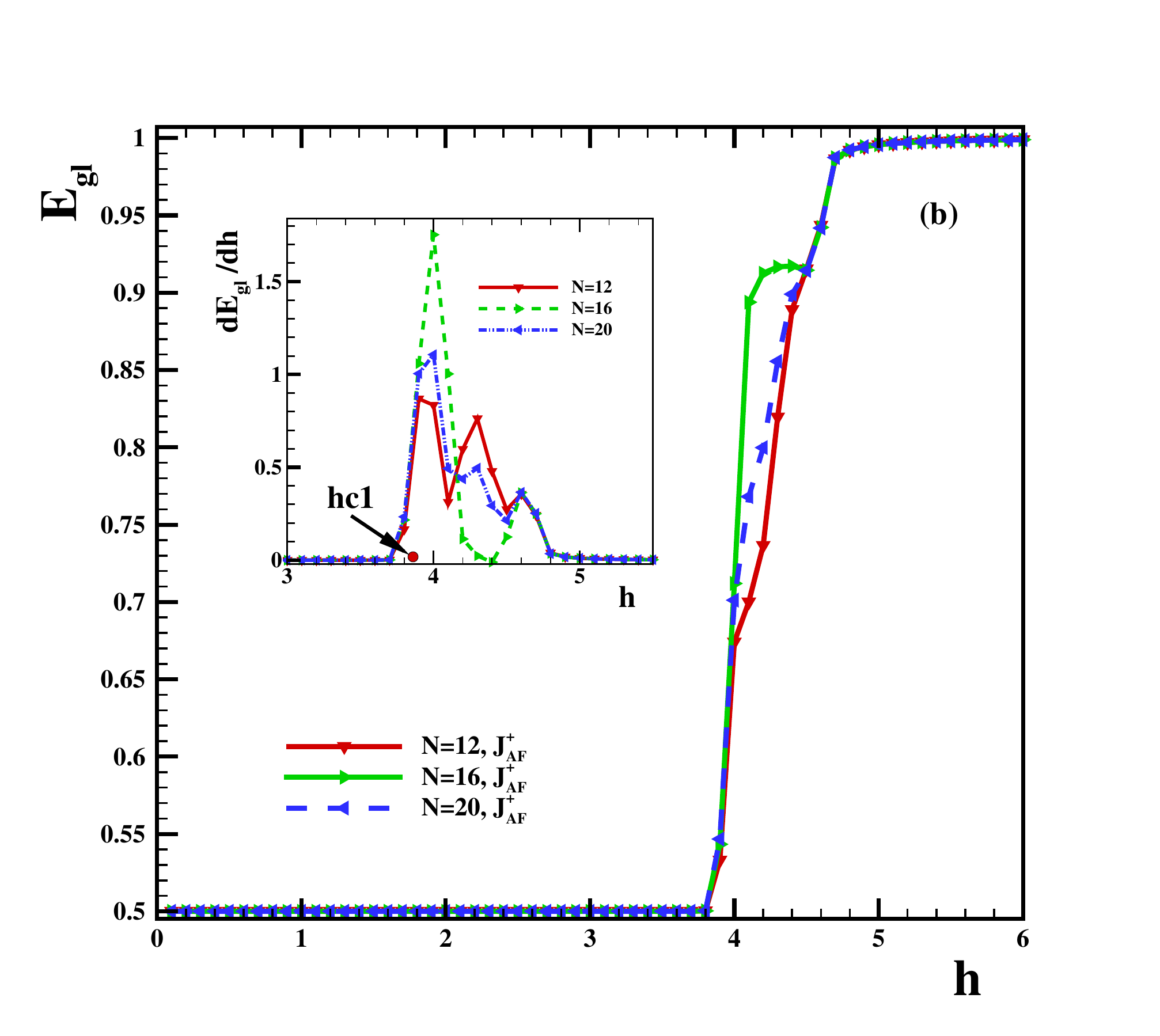}
\caption{(Color online). The global entanglement on (a) weak $AF_{1}$ and (b) strong $AF_{2}$ links respectively, as a function of $h$ for chains with exchanges $J_{F}=1.0,J_{AF}=\frac{9}{2},\delta=\frac{1}{9}$ and lengths N= 12, 16, 20 and $\Delta=0.5$.}
\label{Global}
\end{figure}

\subsection{Global entanglement} \label{secb}
Global-entanglement $(E_{gl})$, offered by Meyer and Wallach \cite{meyer}, measures the per particle total nonlocal information in a general multipartite system\cite{brennen,javad}
\begin{eqnarray}
E_{gl}&=&\frac{1}{N} [2\sum_{i_{1}<i_{2}}\tau_{i_{1}i_{2}}+...+
N \sum_{i_{1}<...<i_{N}}\tau_{i_{1}...i_{N}}],
\label{global entanglement}
\end{eqnarray}
here $E_{gl}$ is the average of tangles scaled with the total number of particles $(\frac{<\tau>}{N})$, without extensive knowledge of tangle distribution among the each particles. Therefore, $E_{gl}$ is an average quantity and cannot recognize between entangled states that have equal tangle  but different distributions of $<\tau>$. To get more intuition, we have computed numerically the Global entanglement between spins. In Fig.~\ref{Global} we have plotted Global entanglement of the same strong, weak bond as a function of the magnetic field for chains with exchanges $J_{F}=1.0, J_{AF}=\frac{9}{2}, \delta=\frac{1}{9}$ and lengths $N= 12, 16, 20$ and $\Delta=0.5$.
\par
The general behavior comes out from $E_{gl} $ is clear that it shows a reverse trend in comparison with the concurrence. Indeed, one can expect such behavior in which for maximal two parties entanglement there is not left any information to share among other parties. So the $E_{gl}$ is exactly zero for the region ($h<h_{c_1}$) which the concurrence is maximum. On the contrary, in the region ($h>h_{c_4}$) where the concurrence is zero, the $E_{gl}$ reaches its maximum value. It signals for the region, which entanglement between two parties is totally absent, the amount of entanglent which is shared between other entities is maximum. However, for $h>h_{c_4}$ system lives in the saturated ferromagnetic phase, but one can see the global entanglement shared on strong and weak antiferromagnetic links is maximum. Moreover, in the  intermediate region for $h_{c_1}<h<h_{c_4}$, the $E_{gl}$ finds value, but its amount is different for weak and strong links. Indeed, by increasing magnetic field the $E_{gl}$ shared among the  weak links $J_{AF}^-$ starts to increase very sharply at $h_{c_4}$ (see the inset of  Fig.~\ref{Global} (a) ) while the $E_{gl}$ shared among the strong links $J_{AF}^+$ starts at $h_{c_1}$ (see the inset of  Fig.~\ref{Global} (b) ).
\section{conclusion}\label{sec-V }
In this paper, we have studied the ground state phase diagram of the antiferromagnetic dominant $(J_{AF}>>J_{F})$ tetrameric spin-$1/2$ chain with anisotropic ferromagnetic coupling in a transverse magnetic field $h$. In the limit where the antiferromagnetic coupling is dominant $J_{AF}>>J_{F}$ we have mapped the model (1), onto  an effective $XYZ$ Heisenberg chain in an external effective field $h^{eff}$. This mapping allowed us to relate the isotropic case $\triangle=1$. Using the accurate Lanczos method of numerically diagonalization for chains up to $N=24$, we have studied the effects of an external magnetic field on the ground state properties of the system. Using the exact diagonalization results, we have calculated the various order parameter as a function of the transverse magnetic field.
\par
In the first part of the numerical experiment, we have investigated the energy gap of the system. In the second part of the numerical experiment, we have studied the magnetization. We also calculate the string correlation function and bond-dimer order parameters. By studying the string correlation function, we found that in the absence of the magnetic field, the suggested alternating chain is in dimer (Haldane) phase and this phase remains stable in the presence external magnetic field up to the first critical field.  In principle, we have addressed five gapped phases in the ground state phase diagram of the system which are separated by four critical fields, $h_{c_{1}}=3.86\pm0.01, h_{c_{2}}=4.01\pm0.01, h_{c_{3}}=4.53\pm0.01, h_{c_{4}}=4.61\pm0.01$.
\par
To get more physical insight we have also investigated the concurrence and global-entanglement between spins as a function of the magnetic field. Both quantities show the occurrence of quantum phase transitions and the corresponding critical points. It is also worth  noting that  the concurrence and global-entanglement show different trend in each mentioned phases.
\vspace{0.3cm}

\end{document}